\documentclass[aps,twocolumn]{revtex4}%
\usepackage{amsfonts}
\usepackage{amsmath}
\usepackage{amssymb}
\usepackage{graphicx}%
\begin{document}
\preprint{ }
\title[Ultra-high-Q small mode volume toroid microcavities]{Demonstration of ultra-high-Q small mode volume toroid microcavities on a chip}
\author{T.J. Kippenberg, S.M. Spillane, K.\ J. Vahala}
\affiliation{Thomas J. Watson Laboratory of Applied Physics, California\ Institute of Technology}
\pacs{PACS number}

\begin{abstract}
Shell document for REV\TeX{} 4.

\end{abstract}
\begin{abstract}
Optical microcavities confine light spatially and temporally and find
application in a wide range of fundamental and applied studies. In many areas,
the microcavity figure of merit is not only determined by photon lifetime (or
the equivalent quality-factor, Q), but also by simultaneous achievement of
small mode volume V . Here we demonstrate ultra-high Q-factor small mode
volume toroid microcavities on-a-chip, which exhibit a Q/V factor of more than
$10^{6}(\lambda/n)^{-3}$. These values are the highest reported to date for
any chip-based microcavity. A corresponding Purcell factor in excess of 200
000 and a cavity finesse of $2.8\times10^{6}$ is achieved, demonstrating that
toroid microcavities are promising candidates for studies of the Purcell
effect, cavity QED or biochemical sensing

\end{abstract}
\volumeyear{2004}
\maketitle

Microcavities can be characterized by two figures of merits: the temporal
confinement is described by the quality factor of the mode Q , and the spatial
confinement is characterized by the mode volume V.\cite{Vahala} Of all optical
microcavities, whispering-gallery-type microsphere resonators have obtained
the highest Q factor to date (nearly 9 billion\cite{Vernooy}\cite{Braginskii}%
). While Q-factor figures prominently in many applications, minimizing mode
volume is also important in a variety of fundamental and applied studies such
as quantum information studies, photonics and biochemical sensing. In
particular, a high Q/V ratio is desirable in applications such as lasers,
add-drop filters and biochemical sensors, which rely upon a large finesse.
Cavities can also be used to enhance the spontaneous emission rate, a concept
which is used in \textquotedblleft single-photon on-demand\textquotedblright%
\ sources,\cite{Moreau}\cite{Santori} where the figure of merit is the Purcell
factor, given by%

\begin{equation}
F=\frac{3}{4\pi^{2}}\frac{Q}{V}\left(  \frac{\lambda}{n}\right)  ^{3}%
\end{equation}

(Ref. \cite{PURCELL}). Wafer-based cavities such as photonic
crystals\cite{Akahane} microposts,\cite{Moreau} or microdisks\cite{Gayral}
typically have much smaller mode volume than microspheres, and allow
wafer-scale integration and control. However, the Q factor of these cavities
has remained significantly lower than for silica microspheres. However,
recently, ultra-high-Q performance on a chip has been demonstrated using a
toroid-microcavity.\cite{Armani} These cavities allow integration and control
previously not accessible in the ultra-high-Q regime. Here we demonstrate that
toroid microcavities not only allow one to obtain ultrahigh- Q, but also to
achieve small mode volumes, reaching a previously inaccessible range of Q/V
ratio of more than $10^{6}(\lambda/n)^{-3}$. By variations of the principal
and minor toroid diameter, the Q/V value was adjusted and the highest achieved
value was$Q/V_{m}=2.5\times10^{6}(\lambda/n)^{-3}$ (for a toroid microcavity
with $\lambda=1550$ $nm$, $V_{m}=180$ $\mu m^{3}$, and $Q_{0}=4\times10^{8}$).
This result is more than one order of magnitude larger than the highest value
reported so far, using crystal defect cavities.\cite{Akahane}\cite{Srinivasan}
Further optimization of the toroid microcavity can result in yet higher
values. The use of silica microtoroids allows the preservation of ultra-high-Q
factors while the additional transverse spatial confinement of the optical
mode over spherical cavities results in a smaller modal volume. To investigate
the modal volumes of toroid microcavities achievable experimentally, samples
were fabricated as described in Ref. \cite{Armani}, by lithography and etching
followed by a CO$_{2}$ laser-assisted reflowprocess. Figure 1 shows a scanning
electron microscopy image of a toroid microcavity. The geometrical parameters
de-fining the structure are the minor toroid diameter d and theprincipal
toroid diameter (denoted $D$). The principal toroid diameter D was controlled
by the size of the initial silica disk preform. The thickness of the toroid
(i.e., minor diameter $d$) is determined by a combination of the initial oxide
thickness and the flux and exposure time of the subsequent CO$_{2}$ laser
anneal. In order to create small mode volume microcavities, 1
%TCIMACRO{\U{b5}}%
%BeginExpansion
$\mu$%
%EndExpansion
m thermally oxidized silicon wafers were used. Subsequently, photolithography
and etching were used to create various diameter silica disk preforms, with
the smallest diameter being 20
%TCIMACRO{\U{b5}}%
%BeginExpansion
$\mu$%
%EndExpansion
m. The laser assisted reflow was used in a final step to create toroidal
microcavities. Duringthis process the silica melts, and the toroid principal
diameter is reduced while silica is consumed to form the toroidal peariphery.
For a 1-%
%TCIMACRO{\U{b5}}%
%BeginExpansion
$\mu$%
%EndExpansion
m-thick thermal oxide, the smallest toroidal cross sections (minor) diameters
obtained in this work were $3.3$ $\mu m$ and principal diameters as small as
$12$ $
\mu
m$ were achieved, as inferred by scanning-electron microscopy. It should be
noted that for oxide layers below 1
$\mu$%
m, it was found that the residual strain in the oxide caused nonuniform silica
disk preforms.

\bigskip
\begin{figure}
[tbp]
\begin{center}
\includegraphics[
height=5.7947cm,
width=7.5278cm
]
{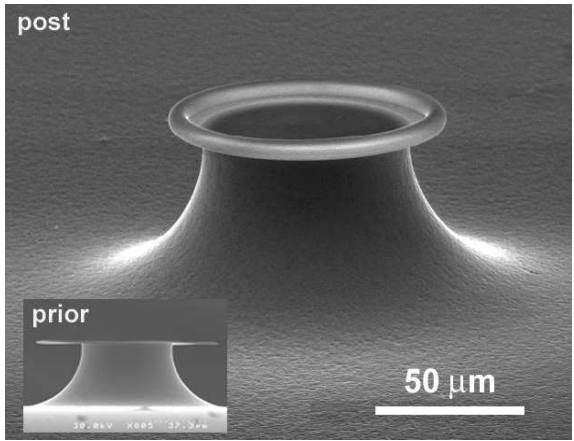}
\caption{Scanning-electron-microscopy image of a high Q/V toroid microcavity
on a chip, using a 1 micron thermal oxide as microdisk as perform (inset). }%
\end{center}
\end{figure}

Characterization of the quality factor of microtoroidsproceeded by coupling
the optical whispering-gallery modes to tapered optical fibers. A
piezoelectric three-axis stage with 20 nm closed-loop resolution allowed
accurate control of the taper-microtoroid coupling gap. Tapers with diameters
of $1-2\mu m$ were used, and efficient coupling with high ideality was
achieved.\cite{Spillane}A narrow line-width external cavity laser was used to
excite the optical modes, and the taper transmission recorded. The
ultra-high-Q toroid modes typically exhibited a doublet structure due to the
presence of scattering centers which can induce modal coupling of the
degenerate clockwise and counterclockwise whispering-gallery
modes.\cite{Weiss}\cite{Kippenberg} The strength of intermode coupling (i.e.,
visibility of the doublet structure, as given by the ratio of the cavity
linewidth $\tau^{-1}/2\pi$ and the splitting frequency $\gamma^{-1}/2\pi$ )
can be described by the dimensionless modal coupling parameter $\Gamma
\equiv\frac{\tau_{0}}{\gamma}$ ,\cite{Kippenberg} and was found to generally
increase for decreasing microcavity size, due to the enhanced
scattering-capture efficiency of the WGM modes. The inset of Fig. 2 shows the
doublet structure of a WGM in a 28-%
%TCIMACRO{\U{b5}}%
%BeginExpansion
$\mu$%
%EndExpansion
m-diameter toroid (with the doublet splitting frequency $\gamma^{-1}%
/2\pi=15.5MHz$, and \ $\Gamma=31$). In addition, the optical modes exhibited
strong thermal effects even at low powers, due to their small mode volume.
These thermal effects induce distortion of the line-shape unless input power
levels are adjusted to a suitably low value. These distortions tend to make
the apparent linewidth larger (pull the resonant frequencies of the mode) in
one frequency scan direction while making its appearance narrower in the
opposing scan direction. These artifacts present a challengeto Q measurement
based upon linewidth measurement alone. In order to accurately determine the Q
in presence of strong modal coupling and thermal effects, an alternative
approach was used. This approach does not rely upon measurement of cavity
linewidth, but on measurement of doublet splitting frequency $\gamma^{-1}%
/2\pi$ and the cavity ringdown time at the critical coupling point.
Alternatively, the cavity Q can be inferred by measurement of the modal
coupling parameter and a ringdown measurement, as is presented in Ref.
\cite{Armani}. In comparison with the latter, the method presented here does
not require knowledge of $\Gamma$ ,\cite{Kippenberg} but rather only the
doublet splitting frequency $\gamma^{-1}/2\pi$. Measurement of splitting
frequency is less sensitive to thermal effects since the splitting frequency
has been observed to be nearly immune to thermal shifts (assuming that each
mode is frequency shifted nearly equally by the excitation wave). Therefore,
the splitting frequency can be deduced even in a regime where the individual
doublets are thermally distorted. To infer the quality factor, the measurement
of the splitting frequency is combinedas in previous studies, with the
ring-down lifetime at the critical coupling point $\tau_{crit}$. The longest
critical decaytime observed in this study was $\tau_{crit}=75$ $ns$. The
following expression is then used to relate both this information and the
measured splitting frequency to the intrinsic Q factor:%

\begin{equation}
Q=\omega\tau_{0}=\omega\frac{\tau_{crit}}{2}\left[  \left(  \frac{1}%
{\tau_{crit}}\right)  ^{2}-\left(  \frac{1}{\gamma}\right)  ^{2}\right]  ^{-1}%
\end{equation}

\bigskip%
\begin{figure}
[tbp]
\begin{center}
\includegraphics[
height=6.5218cm,
width=7.6992cm
]%
{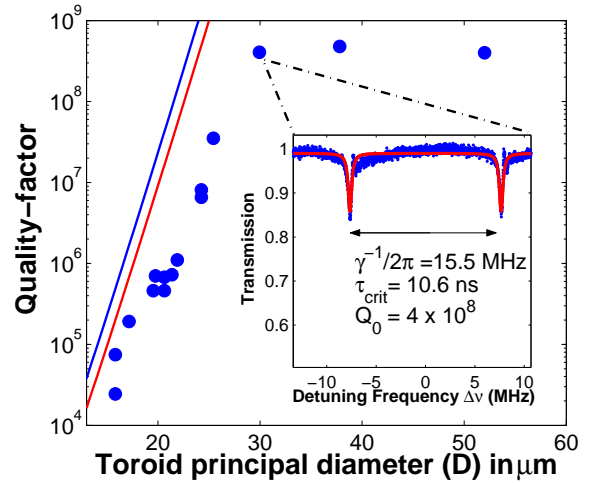}%
\caption{Figure 2: Quality factor and mode volume of toroid microcavities,
with approximately constant minor diameter (d=3.5 $\mu$m) and varying outer
diameter. The solid lines are the whispering-gallery tunnel loss for a
microsphere (theory) and the dots are the measured Q-factors. Inset: The
doublet splitting frequency $\gamma$ and for a $14-\mu m$ radius toroid
microcavity. The critical decay time measured for this sample was 10.6 ns. The
intrinsic Q-factor is 4$\cdot$10$^{8}$, corresponding to a modal coupling
parameter $\Gamma$ of approximately 31.}%
\end{center}
\end{figure}
%EndExpansion

Figure 2 shows the quality factor determined through this method for a number
of microtoroids as a function of toroid principal diameter D for minor
diameter in the range of d$=2.8-6$ $\mu m$ . The highest measured value was 4
108, in a toroid microcavity with geometrical parameters $d=6$ $\mu m,D=29$
$\mu m$, and $\lambda=1550$ $nm$. As evident from the experimental results,
ultra-high-Q (%
%TCIMACRO{\TEXTsymbol{>}}%
%BeginExpansion
$>$%
%EndExpansion
10$^{8})$ factor could be maintained until a limiting diameter of 28
%TCIMACRO{\U{b5}}%
%BeginExpansion
$\mu$%
%EndExpansion
m, below which point a strong decrease of Q factor is observed. Reduction of Q
for smaller principal diameters is expected, as the whispering-gallery loss
increases as a result of the increased optical leakage through the
whispering-gallery potential.\cite{Weinstein}\cite{Datsyuk} In a sphere the
whispering-gallery Q scales as $Q\propto exp(2l)$, where $l$ is the angular
mode number.\cite{Weinstein}\cite{Datsyuk}. Numerical modeling confirms that
the angular mode number for the measured microtoroids is nearly unchanged
compared to that of a microsphere of identical principal diameter, and the
radial eigenfunctions are still well described by their microsphere
counterparts. Therefore, the Q-factor is expected to approximately scale
according to the whispering-gallery loss of a microsphere. For comparison, the
numerically calculated whispering-gallery-loss for a microsphere TE(TM) mode
is also shown in Fig. 2 and track the measured Q toroid Q values in the leaky
regime. Also evident from Fig. 2, the toroid microcavities of this work
nonetheless show Q-values typically more than $l$ order of magnitude below the
expected whispering-gallery-loss limit for a spherical cavity, indicating that
other contributing factors such as surface scattering and surface contaminant
absorption also increase with decreasing diameter. This is consistent with
observations in silica microspheres.\cite{Vernooy} To infer the mode volume of
the fundamental whispering-gallery modes of the microcavities, we used a
numerical finite-element modeling method in conjunction with the exact toroid
geometry (as determined from scanning-electron microscopy) to solve for the
optical modes of the toroid. Mode volume can then be evaluated from the
commonly used definition (where is the dielectric constant $\epsilon$):%

\begin{equation}
V=\frac{\int\epsilon\left\vert \vec{E}\right\vert ^{2}dV}{\max\epsilon
\left\vert \vec{E}\right\vert ^{2}}%
\end{equation}

The inset of Fig. 3 shows the mode volume of the fundamental TE
whispering-gallery modes, as a function of minor toroid diameter $d$ (for a
principal diameter of $25,50,$and $75$ $%
%TCIMACRO{\U{b5}}%
%BeginExpansion
\mu
%EndExpansion
m)$. As expected, the mode volume is continuously reduced as a function of
both decreasing principal diameter $D$ and minor diameter $d$ . For a fixed
principal diameter $D$, reduction of minor diameter causes first a slow
reduction of mode volume [scaling as $\left(  d/D\right)  ^{1/4}$, due to
azimuthal compression of the mode], followed by a strong compression of the
modein both radial and azimuthal directions. Figure 3 shows the $Q/V$ ratio
determined for the samples under consideration in this study. The $Q/V$ value
exhibits an optimum, which occurs near a principal diameter of $29$ $\mu m$.
At large principal diameters the intrinsic quality factor is experimentally
found to be nearly independent of radius (believed to be limited byscattering
and OH absorption contributions) and for this reason reduction of the
principal cavity diameter allows an increase in $Q/V$ ratio. However, the
$Q/V$ value cannot be increased indefinitely, as the reduction of mode volume
ultimately comes at the expense of reduced quality factor due to the presence
of tunneling loss. For radii where the tunnel loss becomes the dominant loss
mechanism, the reduction in mode volume is more than offset by a much more
strongly reduced quality factor, adversely affecting the $Q/V$ value.
Therefore, for a given minor diameter \thinspace\ , an optimum $Q/V$ value
exists. More generally, since the mode volume (and Q) depends on both minor
and outer toroid diameter, there is also an absolute maximum of the $Q/V$
value for toroid microcavities. Numerical modeling of the maximum achievable
$Q/V$ for a toroidal geometry was also investigated, and will be presented elsewhere.

\bigskip
\begin{figure}
[tbp]
\begin{center}
\includegraphics[
height=2.22in,
width=2.738in
]%
{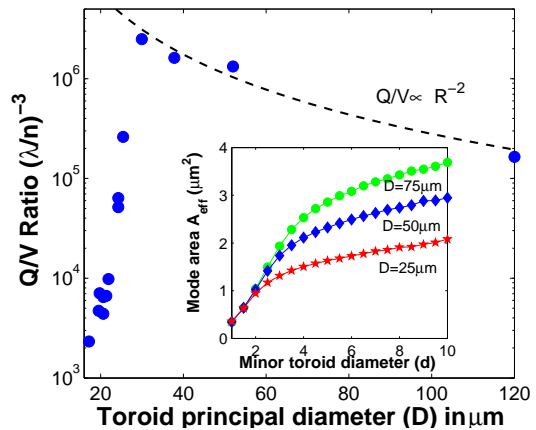}%
\caption{The ratio of Q-factor and mode volume (V) of the experimental data
from figure 2. The mode volume was estimated by numerical finite element
modeling using the cavity geometry parameters, as inferred by SEM. Inset: The
mode area of a toroid microcavity as a function of cross sectional diameter
(d) for fixed principal toroid diameter (D=25,50,75 $\mu$m are shown). The
highest Q/V value of more than 106($\lambda$/n)-3 was measured for a 28-$\mu
$m-diameter toroid.}%
\end{center}
\end{figure}
%EndExpansion

The maximum achievable Q/V factor in these studies was $2.5\times10^{6}\left(
\lambda/n\right)  ^{-3}$, for a microtoroid with $D=29\mu m,d=6\mu
m,Q=4\times10^{8},V_{m}=180$ $\mu m^{3}$ at $\lambda=1550$ $nm$. The highest
reported Q/V factor to date for an optical microcavity is $4.5\cdot
10^{4}\left(  \lambda/n\right)  ^{-3}$, for a photonic crystal defect
cavity.\cite{Akahane} Therefore, the measured value in this work constitutes a
more than one order of magnitude improvement, and enters an unprecedented
$Q/V$ range of more than $10^{6}\left(  \lambda/n\right)  ^{-3}$. To allow
further comparison with other microcavity geometries, the Purcell \ factor and
finesse are also calculated. The cavity-finesse $2.8=10^{6}$ exceeds the
highest value reported so far of $2.3=10^{6}$.\cite{Rempe} In addition a
Purcell-factor of $2\cdot10^{5}$ is achieved, which is an improvement of more
than one order of magnitude over previously reported values\cite{Akahane}%
\cite{Srinivasan}. In summary, ultra-high-Q small mode volume toroid
microcavities are demonstrated. By reduction of the toroidal cross section and
variation of the cavity principal diameteran optimum ratio of quality factor
to mode volume $Q/V$ was obtained. The mode volume reduction was
calculatedusing numerical modeling of the toroid modes. The highest achieved
$Q/V$ value was in excess of $10^{6}\left(  \lambda/n\right)  ^{-3}$ and is
more than one order of magnitude larger than for any other chip based
microcavity. Corresponding Purcell factor and cavity finesse are also record
values for in this device. The large Purcell-factor shows that toroid
microcavities are promising candidates for achieving enhancement of
spontaneous emission, and related studies.

\subsection{Acknowledgements}
This work was funded by the DARPA, NSF, and the Caltech Lee Center for
Advanced Networking.


\begin{thebibliography}{15}
\expandafter\ifx\csname natexlab\endcsname\relax\def\natexlab#1{#1}\fi
\expandafter\ifx\csname bibnamefont\endcsname\relax
\def\bibnamefont#1{#1}\fi
\expandafter\ifx\csname bibfnamefont\endcsname\relax
\def\bibfnamefont#1{#1}\fi
\expandafter\ifx\csname citenamefont\endcsname\relax
\def\citenamefont#1{#1}\fi
\expandafter\ifx\csname url\endcsname\relax
\def\url#1{\texttt{#1}}\fi
\expandafter\ifx\csname urlprefix\endcsname\relax\def\urlprefix{URL }\fi
\providecommand{\bibinfo}[2]{#2}
\providecommand{\eprint}[2][]{\url{#2}}
\bibitem[{\citenamefont{Vahala}(2003)}]{Vahala}
\bibinfo{author}{\bibfnamefont{K.~J.} \bibnamefont{Vahala}},
\bibinfo{journal}{Nature} \textbf{\bibinfo{volume}{424}},
\bibinfo{pages}{839} (\bibinfo{year}{2003}).
\bibitem[{\citenamefont{Vernooy et~al.}(1998)\citenamefont{Vernooy, Ilchenko,
Mabuchi, Streed, and Kimble}}]{Vernooy}
\bibinfo{author}{\bibfnamefont{D.~W.} \bibnamefont{Vernooy}},
\bibinfo{author}{\bibfnamefont{V.~S.} \bibnamefont{Ilchenko}},
\bibinfo{author}{\bibfnamefont{H.}~\bibnamefont{Mabuchi}},
\bibinfo{author}{\bibfnamefont{E.~W.} \bibnamefont{Streed}},
\bibnamefont{and} \bibinfo{author}{\bibfnamefont{H.~J.}
\bibnamefont{Kimble}}, \bibinfo{journal}{Optics Letters}
\textbf{\bibinfo{volume}{23}}, \bibinfo{pages}{247} (\bibinfo{year}{1998}).
\bibitem[{\citenamefont{Braginskii et~al.}(1990)\citenamefont{Braginskii,
Ilchenko, and Gorodetskii}}]{Braginskii}
\bibinfo{author}{\bibfnamefont{V.~B.} \bibnamefont{Braginskii}},
\bibinfo{author}{\bibfnamefont{V.~S.} \bibnamefont{Ilchenko}},
\bibnamefont{and} \bibinfo{author}{\bibfnamefont{M.~L.}
\bibnamefont{Gorodetskii}}, \bibinfo{journal}{Uspekhi Fizicheskikh Nauk}
\textbf{\bibinfo{volume}{160}}, \bibinfo{pages}{157} (\bibinfo{year}{1990}).
\bibitem[{\citenamefont{Moreau et~al.}(2001)\citenamefont{Moreau, Robert,
Gerard, Abram, Manin, and Thierry-Mieg}}]{Moreau}
\bibinfo{author}{\bibfnamefont{E.}~\bibnamefont{Moreau}},
\bibinfo{author}{\bibfnamefont{I.}~\bibnamefont{Robert}},
\bibinfo{author}{\bibfnamefont{J.~M.} \bibnamefont{Gerard}},
\bibinfo{author}{\bibfnamefont{I.}~\bibnamefont{Abram}},
\bibinfo{author}{\bibfnamefont{L.}~\bibnamefont{Manin}}, \bibnamefont{and}
\bibinfo{author}{\bibfnamefont{V.}~\bibnamefont{Thierry-Mieg}},
\bibinfo{journal}{Applied Physics Letters} \textbf{\bibinfo{volume}{79}},
\bibinfo{pages}{2865} (\bibinfo{year}{2001}).
\bibitem[{\citenamefont{Santori et~al.}(2002)\citenamefont{Santori, Fattal,
Vuckovic, Solomon, and Yamamoto}}]{Santori}
\bibinfo{author}{\bibfnamefont{C.}~\bibnamefont{Santori}},
\bibinfo{author}{\bibfnamefont{D.}~\bibnamefont{Fattal}},
\bibinfo{author}{\bibfnamefont{J.}~\bibnamefont{Vuckovic}},
\bibinfo{author}{\bibfnamefont{G.~S.} \bibnamefont{Solomon}},
\bibnamefont{and} \bibinfo{author}{\bibfnamefont{Y.}~\bibnamefont{Yamamoto}},
\bibinfo{journal}{Nature} \textbf{\bibinfo{volume}{419}},
\bibinfo{pages}{594} (\bibinfo{year}{2002}).
\bibitem[{\citenamefont{EM}(1946)}]{PURCELL}
\bibinfo{author}{\bibfnamefont{P.}~\bibnamefont{EM}},
\bibinfo{journal}{Physical Review} \textbf{\bibinfo{volume}{69}},
\bibinfo{pages}{681} (\bibinfo{year}{1946}).
\bibitem[{\citenamefont{Akahane et~al.}(2003)\citenamefont{Akahane, Asano,
Song, and Noda}}]{Akahane}
\bibinfo{author}{\bibfnamefont{Y.}~\bibnamefont{Akahane}},
\bibinfo{author}{\bibfnamefont{T.}~\bibnamefont{Asano}},
\bibinfo{author}{\bibfnamefont{B.~S.} \bibnamefont{Song}}, \bibnamefont{and}
\bibinfo{author}{\bibfnamefont{S.}~\bibnamefont{Noda}},
\bibinfo{journal}{Nature} \textbf{\bibinfo{volume}{425}},
\bibinfo{pages}{944} (\bibinfo{year}{2003}), \bibinfo{note}{737KY NATURE}.
\bibitem[{\citenamefont{Gayral et~al.}(1999)\citenamefont{Gayral, Gerard,
Lemaitre, Dupuis, Manin, and Pelouard}}]{Gayral}
\bibinfo{author}{\bibfnamefont{B.}~\bibnamefont{Gayral}},
\bibinfo{author}{\bibfnamefont{J.~M.} \bibnamefont{Gerard}},
\bibinfo{author}{\bibfnamefont{A.}~\bibnamefont{Lemaitre}},
\bibinfo{author}{\bibfnamefont{C.}~\bibnamefont{Dupuis}},
\bibinfo{author}{\bibfnamefont{L.}~\bibnamefont{Manin}}, \bibnamefont{and}
\bibinfo{author}{\bibfnamefont{J.~L.} \bibnamefont{Pelouard}},
\bibinfo{journal}{Applied Physics Letters} \textbf{\bibinfo{volume}{75}},
\bibinfo{pages}{1908} (\bibinfo{year}{1999}).
\bibitem[{\citenamefont{Armani et~al.}(2003)\citenamefont{Armani, Kippenberg,
Spillane, and Vahala}}]{Armani}
\bibinfo{author}{\bibfnamefont{D.~K.} \bibnamefont{Armani}},
\bibinfo{author}{\bibfnamefont{T.~J.} \bibnamefont{Kippenberg}},
\bibinfo{author}{\bibfnamefont{S.~M.} \bibnamefont{Spillane}},
\bibnamefont{and} \bibinfo{author}{\bibfnamefont{K.~J.}
\bibnamefont{Vahala}}, \bibinfo{journal}{Nature}
\textbf{\bibinfo{volume}{421}}, \bibinfo{pages}{925} (\bibinfo{year}{2003}).
\bibitem[{\citenamefont{Srinivasan et~al.}(2004)\citenamefont{Srinivasan,
Barclay, Borselli, and Painter}}]{Srinivasan}
\bibinfo{author}{\bibfnamefont{K.}~\bibnamefont{Srinivasan}},
\bibinfo{author}{\bibfnamefont{P.~E.} \bibnamefont{Barclay}},
\bibinfo{author}{\bibfnamefont{M.}~\bibnamefont{Borselli}}, \bibnamefont{and}
\bibinfo{author}{\bibfnamefont{O.}~\bibnamefont{Painter}},
\bibinfo{journal}{ArXiv}  (\bibinfo{year}{2004}).
\bibitem[{\citenamefont{Spillane et~al.}(2003)\citenamefont{Spillane,
Kippenberg, Painter, and Vahala}}]{Spillane}
\bibinfo{author}{\bibfnamefont{S.~M.} \bibnamefont{Spillane}},
\bibinfo{author}{\bibfnamefont{T.~J.} \bibnamefont{Kippenberg}},
\bibinfo{author}{\bibfnamefont{O.~J.} \bibnamefont{Painter}},
\bibnamefont{and} \bibinfo{author}{\bibfnamefont{K.~J.}
\bibnamefont{Vahala}}, \bibinfo{journal}{Physical Review Letters}
\textbf{\bibinfo{volume}{91}}, \bibinfo{pages}{art. no.}
(\bibinfo{year}{2003}).
\bibitem[{\citenamefont{Kippenberg et~al.}(2002)\citenamefont{Kippenberg,
Spillane, and Vahala}}]{Kippenberg}
\bibinfo{author}{\bibfnamefont{T.~J.} \bibnamefont{Kippenberg}},
\bibinfo{author}{\bibfnamefont{S.~M.} \bibnamefont{Spillane}},
\bibnamefont{and} \bibinfo{author}{\bibfnamefont{K.~J.}
\bibnamefont{Vahala}}, \bibinfo{journal}{Optics Letters}
\textbf{\bibinfo{volume}{27}}, \bibinfo{pages}{1669} (\bibinfo{year}{2002}).
\bibitem[{\citenamefont{Weinstein}(1696)}]{Weinstein}
\bibinfo{author}{\bibfnamefont{L.}~\bibnamefont{Weinstein}},
\emph{\bibinfo{title}{Open Resonators and Open Waveguides}}
(\bibinfo{publisher}{The Golem Press}, \bibinfo{address}{Boulder, Colorado},
\bibinfo{year}{1696}).
\bibitem[{\citenamefont{Datsyuk}(1992)}]{Datsyuk}
\bibinfo{author}{\bibfnamefont{V.~V.} \bibnamefont{Datsyuk}},
\bibinfo{journal}{Applied Physics B-Photophysics and Laser Chemistry}
\textbf{\bibinfo{volume}{54}}, \bibinfo{pages}{184} (\bibinfo{year}{1992}).
\bibitem[{\citenamefont{Rempe et~al.}(1992)\citenamefont{Rempe, Thompson,
Kimble, and Lalezari}}]{Rempe}
\bibinfo{author}{\bibfnamefont{G.}~\bibnamefont{Rempe}},
\bibinfo{author}{\bibfnamefont{R.~J.} \bibnamefont{Thompson}},
\bibinfo{author}{\bibfnamefont{H.~J.} \bibnamefont{Kimble}},
\bibnamefont{and} \bibinfo{author}{\bibfnamefont{R.}~\bibnamefont{Lalezari}},
\bibinfo{journal}{Optics Letters} \textbf{\bibinfo{volume}{17}},
\bibinfo{pages}{363} (\bibinfo{year}{1992}).
\end{thebibliography}
\end{document}